\documentclass[11pt]{article}
\usepackage{geometry}
\usepackage{graphicx}
\usepackage{latexsym}
\usepackage{amsmath,amsfonts,amsthm,bm}
\usepackage{epstopdf}
\usepackage[font=small, labelfont=bf, width=1.0\textwidth]{caption}

\def\be{\begin{equation}}
\def\ee{\end{equation}}

\title{A numerical adaptation of SAW identities from the honeycomb to other 2D lattices}
\author{Nicholas R Beaton, Anthony J Guttmann and Iwan Jensen \\ \ \\ {\small ARC Centre of Excellence for Mathematics and Statistics of Complex Systems} \\ {\small Department of Mathematics and Statistics} \\ {\small The University of Melbourne, VIC 3010, Australia}}
\begin{document}
\maketitle
\abstract{
Recently, Duminil-Copin and Smirnov proved a long-standing conjecture by Nienhuis
that the connective constant of self-avoiding walks on the honeycomb lattice is $\sqrt{2+\sqrt{2}}.$ 
A key identity used in that proof depends on the existence of a parafermionic observable for 
self-avoiding walks on the honeycomb lattice. 
Despite the absence of a corresponding observable for SAW on the square and triangular lattices, 
we show that in the limit of large lattices, some of the consequences observed on the honeycomb lattice 
persist on other lattices. This permits the accurate estimation, though not an exact evaluation,
of certain critical amplitudes, as well as critical points, for these lattices. For the honeycomb lattice an exact amplitude for loops is proved. \let\thefootnote\relax\footnotetext{Email: \texttt{nbeaton, t.guttmann, i.jensen@ms.unimelb.edu.au}}
}

\section{Introduction}

\begin{figure}[h]
\begin{center}
\includegraphics[height=200pt]{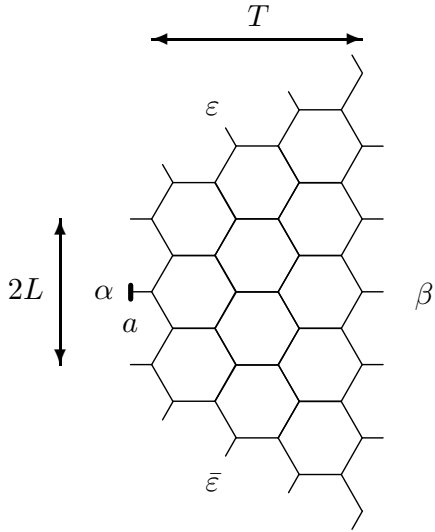}
\end{center}
\caption{Finite patch $S_{3,1}$ of the honeycomb lattice. Paths run from mid-edge to mid-edge acquiring a 
weight $z$ for each step. The paths start on the central mid-edge of the left boundary (shown as $a$). 
}
\label{fig:trap}
\end{figure}

In 2010  Duminil-Copin and Smirnov~\cite {DC-S10} proved that the critical point of honeycomb lattice SAW 
is $z_c = 1/ \sqrt{2+\sqrt{2}}$, as conjectured by Nienhuis~\cite{N82} in 1982. Their proof rested 
on establishing a connection between three  generating functions for walks in a regular trapezoidal 
sublattice $S_{T,L}$ of the honeycomb lattice, as shown in Figure \ref{fig:trap}, of width $T$ 
and left-height $2L.$ All walks start on the half-edge $a$ incident on the left wall labelled $\alpha.$ 

They first identified a parafermionic observable: 
$$F(x) = \sum_{\omega \subset S: \, a \to x} e^{-i\sigma W_\omega(a,x) } z^{l(\omega)}.$$
Here $x$ is a point (specifically, the mid-point between two adjacent vertices) in the domain, $\sigma \in {\mathbb R}$, $z \ge 0$, and $\omega$ is a path starting at 
half-edge $a$ and finishing at point $x.$ $l(\omega)$ is the length of the path and $W_\omega(a,x)$ 
is the winding, or total rotation in radians, when $\omega$ is traversed from $a$ to $x.$

In the special case $z = z_c = 1/\sqrt{2+ \sqrt{2}}$ and $\sigma = 5/8,$ $F$  satisfies half of the discrete Cauchy-Riemann equations \cite{Smirnov10},
\begin{equation}\label{eqn:local_identity}
(p-v)F(p) + (y-v)F(q) + (r-v)F(r) = 0,
\end{equation} 
where $p,$ $q,$ $r$ are the mid-edges of  the three edges incident on vertex $v.$

By summing \eqref{eqn:local_identity} over all vertices in $S_{T,L}$, contributions from interior mid-edges cancel and we are left with a relation involving only mid-edges on the boundary of $S_{T,L}$. Walks from half-edge $a$ to the boundary fall into three classes, depending on which boundary edge they terminate at. 
This observation gives rise to three generating functions:

\begin{align*}
A_{T,L}(z) &:= \sum_{\genfrac{}{}{0pt}{}{\gamma\subset S_{T,L}}{a\rightarrow \alpha\slash\{a\}}} z^{\ell(\gamma)},\\
B_{T,L}(z) &:= \sum_{\genfrac{}{}{0pt}{}{\gamma\subset S_{T,L}}{a\rightarrow \beta}} z^{\ell(\gamma)},\\
%\hat{B}_{T,L}(x,y) &:= \sum_{\genfrac{}{}{0pt}{}{\gamma\subset S_{T,L}}{a\rightarrow \beta, W_{\gamma}=\pm \pi/3}} x^{-\ell(\gamma)}y^{\nu(\gamma)},\\
E_{T,L}(z) &:= \sum_{\genfrac{}{}{0pt}{}{\gamma\subset S_{T,L}}{a\rightarrow \varepsilon\cup\bar{\varepsilon}}} z^{\ell(\gamma)},
\end{align*}
where the sums are over all self-avoiding walks from $a$ to the $\alpha$, $\beta$ or $\varepsilon,\bar{\varepsilon}$  
boundaries, respectively. 

Duminil-Copin and Smirnov showed that the relation involving boundary mid-edges can be written in the form

\be \label{eqn:li}
1=\cos \left(\frac{3\pi}{8}\right) A_{T,L}(z_{\rm c}) + \cos \left(\frac{\pi}{4}\right) E_{T,L}(z_{\rm c}) +  B_{T,L}(z_{\rm c}).
\ee
Taking the limit $L \to \infty$, this becomes
\be \label{eqn:li2}
1=\cos \left(\frac{3\pi}{8}\right) A_{T}(z_{\rm c}) + \cos \left(\frac{\pi}{4}\right) E_{T}(z_{\rm c}) +  B_{T}(z_{\rm c}).
\ee
This is now a relation linking three generating functions for SAW in a strip of width $T.$ Duminil-Copin and Smirnov 
then used this to prove Nienhuis's conjecture. 

In \cite{BdGG11} it was proved that $E_{T}(z_{\rm c}) = 0,$  so (\ref{eqn:li2}) can in fact be simplified to 

\be \label{eqn:li3}
1=\cos \left(\frac{3\pi}{8}\right) A_{T}(z_{\rm c}) +  B_{T}(z_{\rm c}).
\ee
This modification slightly simplifies the proof of Duminil-Copin and Smirnov. 

This simpler identity, involving just two generating functions, raises the question as to its applicability to 
other two dimensional lattices, such as the square and triangular lattices. However, it should be 
remarked that the critical points are not known exactly for these lattices, though high precision numerical 
estimates are available. The proof of Duminil-Copin and Smirnov identifying the critical point cannot be 
repeated for these lattices, as there is no known appropriate parafermionic observable satisfying 
an identity like~\eqref{eqn:local_identity}. As shown by Ikhlef and Cardy~\cite{CI09}, the dilute O$(n)$ model on the square lattice does have a 
parafermionic observable, and this can be used to identify its critical point. In the $n \to 0$ limit, however, that model is not the usual SAW model. 
For the triangular lattice we know of no appropriate parafermionic observable.

However, leaving aside considerations of discrete holomorphicity and the existence of parafermions, 
and motivated by a more detailed study of the width ($T$) dependence of the two generating functions 
$A_{T}(z)$ and  $B_{T}(z),$ we investigated which features of the identity (\ref{eqn:li3}) carry over to 
other two-dimensional lattices. 

\begin{figure}[h!]
\begin{center}
\includegraphics[height=250pt]{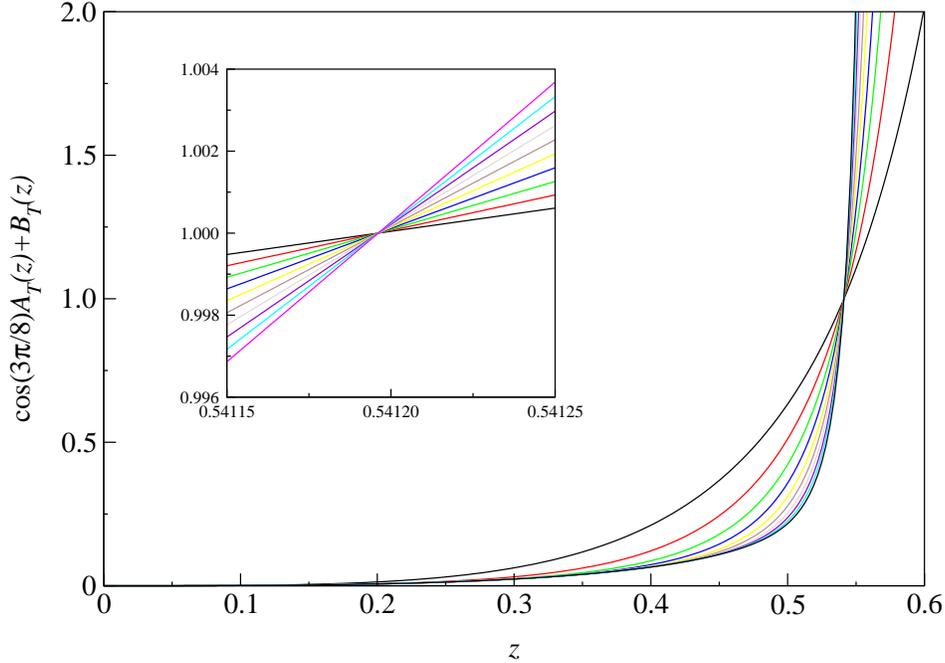}
\end{center}
\caption{Plot of  $\cos(3\pi/8)A_{T}(z) + B_{T}(z)$ for $T \in [1..10],$ for the honeycomb lattice, 
showing interception of all curves at $(z_c,1).$ The insert shows a close-up or the region of interception. }
\label{hcAB}
\end{figure}

We did this by calculating data for the generating functions $A_{T}(z)$ and  $B_{T}(z)$ in strips, 
for $T \le 10$ on the honeycomb lattice, for $T \le 15$ on the square lattice and for $T \le 11$ on the triangular lattice. 
In Figure \ref{hcAB} we show a plot of  $\cos(3\pi/8)A_{T}(z) + B_{T}(z)$ for $T \in [1\dots10],$ on the honeycomb lattice, 
showing the intersection of all curves at $(z_c,1),$ as expected from (\ref{eqn:li3}).

When we repeated this calculation using the square and triangular lattice data, we found that there was no 
unique point of intersection, though the discrepancy was too small to be visible on a plot such as Figure~\ref{hcAB}. 
Instead we observed a definite  monotonic width dependence, described in greater detail below. So we first 
investigated a more general form of (\ref{eqn:li3}), allowing the ``constants" to be width dependent. That is to say, we assumed
$$1=c_\alpha(T)A_{T}(z_{\rm c}) +  c_\beta(T) B_{T}(z_{\rm c}).$$ 
We fitted successive pairs of points $(A_{T}(z_{\rm c}), B_{T}(z_{\rm c}))$ and $(A_{T+1}(z_{\rm c}), B_{T+1}(z_{\rm c}))$ 
in order to estimate $c_\alpha(T)$ and $ c_\beta(T),$ and found a weak $T$ dependence in both ``constants''.
More significantly however, we conjecture that 
$$\lim_{T \to \infty} c_\alpha(T)/c_\beta(T) = \cos(3\pi/8),$$ 
just as in the honeycomb lattice case, based on agreement to 
more than 5 significant digits for both the square and triangular lattices. In hindsight, this is perhaps not too 
surprising, as the constants multiplying the two generating functions arise from the winding angle of 
contributing graphs, and these are independent of lattice for the two generating functions considered, 
being $\pm\pi$ radians for $A(z)$ and $0$ for $B(z).$

Assuming this ratio is indeed $\cos(3\pi/8)$ for all lattices, we re-analysed the data with this assumption implicit. 
That is to say, we fitted the data to 
$$c(T)=\cos(3\pi/8) A_{T}(z_{\rm c}) +  B_{T}(z_{\rm c}).$$ 
We were able to estimate both the limit $\lim_{T \to \infty} c(T),$ which  is of course  
lattice dependent, and also the nature of the $T$ dependence. Putting all this together we estimated for the square lattice, 
\begin{equation} \label{eq:ki4sq}
1.024966(1 - 0.14/T^2)\approx\cos(3\pi/8) A_{T}(z_{\rm c}) +  B_{T}(z_{\rm c}).
\end{equation}
  For the triangular lattice the corresponding result is
  \begin{equation} \label{eq:ki4tr}
1.901979(1 - 0.1/T^2)\approx\cos(3\pi/8) A_{T}(z_{\rm c}) +  B_{T}(z_{\rm c}).
\end{equation}
The leading constant is expected to be sufficiently accurate so as to restrict errors to 1 or 2 places 
in the last quoted digit. The correction term is given as O$(1/T^2)$ but that exponent is a guess based 
an a numerical estimate in the range $(1.9,2.1).$ Finally the magnitude of that term is  claimed to be 
accurate only to $10\% \,\, {\rm to} \,\, 20\%.$

This limiting behaviour as $T \to \infty$ suggests a new numerical method for estimating the critical point. 
For the honeycomb lattice the intersection point of $\cos(3\pi/8)A_{T}(z) + B_{T}(z)$ for any two distinct 
values of $T$ uniquely determines $z_c.$ For the square and triangular lattices, we instead looked at the 
intersection point of $\cos(3\pi/8)A_{T}(z) + B_{T}(z)$ and $\cos(3\pi/8)A_{T+1}(z) + B_{T+1}(z).$ 
Call this intersection point $z_c(T).$ Then one expects $\lim_{T \to \infty} z_c(T) = z_c.$ 

In this way we estimated 
$$z_c = 0.3790522775 \pm 0.0000000005$$ 
for the square lattice and 
$$z_c = 0.240917572 \pm 0.000000005$$ 
for the triangular lattice.  These estimates can be compared to the best series estimates, 
based on analysis of very long polygon series, which are  $z_c({\rm sq})=0.37905227776$  \cite{JG99,IJ03},  
with uncertainty in the last digit, and  $z_c({\rm tr}) = 0.2409175745$  \cite{IJ04b}, with similar uncertainty.

Another result proved by Beaton, de Gier and Guttmann \cite{BdGG11} is that $\lim_{T \to \infty} B_T(z) = B(z) = 0$ for $z \le z_c.$ 
This means that, in this limit, the identity (\ref{eqn:li3}) further simplifies to \begin{equation}\label{eq:ki4}
1 = \cos(3\pi/8)A(z_c), \end{equation} where $A(z) = \lim_{T \to \infty} A_T(z).$ 
Now $A(z)$ is the generating function for loops (half-plane walks which begin and end on the boundary), which is expected to behave \cite{GT84} as 
$$A(z) \sim a_0 + a_1(1 - z/z_c)^{3/16},$$ 
so~\eqref{eq:ki4} implies an {\em exact} value for the critical amplitude:
$$a_0 = 1/\cos(3\pi/8).$$ 
Correspondingly, highly accurate predictions for this amplitude for the square and triangular lattices 
follow from ~(\ref{eq:ki4sq}) and (\ref{eq:ki4tr}), notably
$$a_0({\rm sq}) \approx 1.024966/\cos(3\pi/8)=2.678365$$ 
and $$a_0({\rm tr}) \approx 1.901979 /\cos(3\pi/8) = 4.970111.$$
We remark that, as discussed below, we have normalised these generating functions differently 
for these two lattices, so please note details of the normalization before using these estimates.

Finally, we studied the behaviour of the two generating functions as $T \to \infty.$ We found
$B_T(z) \sim {\rm const.}/T^\alpha,$ where $\alpha \approx 0.25.$  This is precisely as predicted in  \cite{DC-S10}, 
based on conjectures of Lawler, Schramm and Werner  \cite{LSW04} as to the number of SAWs on the 
boundary of a domain. In \cite{DC-S10} it was pointed out that the conjecture implies that $B_T(z_c)$ 
should decay as $T^{-1/4}$ as $T$ goes to infinity, just as we observed. 

Similarly, we can investigate how $\tilde {A}_T(z_c) =  A_T(z_c) - a_0 $ decays as $T$ tends to infinity. 
From~\eqref{eqn:li3} it follows that  ${\tilde A}_T(z_c)$ also decays like $T^{-1/4}$ and this was observed numerically.

\section{Honeycomb lattice} \label{honey}
The original identity of Duminil-Copin and Smirnov related three distinct 
generating functions for SAW in a finite domain. Letting the length $L$ of the domain become unbounded 
changes the domain into a strip of finite width $T.$ As  proved in \cite{BdGG11}, the generating function 
$E_T(z)$ vanishes identically in that limit, so one has an identity relating two generating functions:
\begin{equation} \label{keyh}
1=\cos(3\pi/8)A_{T}(z_c) + B_{T}(z_c).
\end{equation}

 The generating functions $A_T(z)$ and $B_T(z)$ for $T$ finite are rational. For example, 
$$A_0(z) = \frac{2 z^3}{1-z^2}, \,\,\,\,
B_0(z) = \frac{2 z^2}{1-z^2}.$$

$$A_1(z) = \frac{2 z^3 \left(1-z^2+z^4+3 z^6-4 z^8+z^{12}\right)}{\left(1-z^4\right)^2 \left(1-2 z^2+z^4-z^6\right)},$$
$$B_1(z) = \frac{2 z^4 \left(2-4 z^4+2 z^6+2 z^8-z^{10}\right)}{\left(1-z^4\right)^2 \left(1-2 z^2+z^4-z^6\right)}.$$

$$A_2(z) = \frac{P_2^{\rm A}(z)}{Q_2(z)}, \,\,\,\,
B_2(z) = \frac{P_2^{\rm B}(z)}{Q_2(z)}, \qquad \mbox{where}$$

\begin{eqnarray*}
P_2^{\rm A}(z) & = & 2 \left(z^3\!-\!4 z^5\!+\!7 z^7\!-\!7 z^9\!+\!9 z^{11}\!+\!2 z^{13}\!-\!31 z^{15}\!+\!39 z^{17}\!-\!46 z^{19}\!+\!68 z^{21}\! \right.\\
& & \qquad \left. -75 z^{23}\!+\!74 z^{25}\!-\!61 z^{27}\!+\!41 z^{29}\!-\!20 z^{31}\!+\!z^{33}\!+\!6 z^{35}\!-\!4 z^{37}\!+\!z^{39}\right) \\ \\
 P_2^{\rm B}(z) & =& 2 z^6 \left(1\!-\!z^2\right) \left(4\!-\!4 z^2\!-\!8 z^4\!+\!8 z^6\!-\!4 z^8\!+\!16 z^{10}\!-\!12 z^{12}\!+\!18 z^{14}\!\!\right.\\
& & \qquad  \qquad \qquad \left.  -10 z^{16}\!+\!3 z^{18}\!-\!3 z^{20}\!-\!4 z^{22}\!+\!10 z^{24}\!-\!10 z^{26}\!+\!5 z^{28}\!-\!z^{30}\right)\\ \\
Q_2(z) & = & \left(1\!-\!z^2\!-\!z^4\!+\!z^6\!-\!z^8\right)^2 \times \\ 
& & \left(1\!-\!3 z^2\!+\!3 z^4\!-\!5 z^6\!+\!8 z^8\!-\!9 z^{10}\!+\!7 z^{12}\!-\!8 z^{14}\!+\!8 z^{16}\!-\!5 z^{18}\!+\!3 z^{20}\!-\!z^{22}\right)\\
\end{eqnarray*}
%
%\tiny{
%$$A_2(z) = \frac{2 \left(z^3\!-\!4 z^5\!+\!7 z^7\!-\!7 z^9\!+\!9 z^{11}\!+\!2 z^{13}\!-\!31 z^{15}\!+\!39 z^{17}\!-\!46 z^{19}\!+\!68 z^{21}\!-\!75 z^{23}\!+\!74 z^{25}\!-\!61 z^{27}\!+\!41 z^{29}\!-\!20 z^{31}\!+\!z^{33}\!+\!6 z^{35}\!-\!4 z^{37}\!+\!z^{39}\right)}{\left(1\!-\!z^2\!-\!z^4\!+\!z^6\!-\!z^8\right)^2 \left(1\!-\!3 z^2\!+\!3 z^4\!-\!5 z^6\!+\!8 z^8\!-\!9 z^{10}\!+\!7 z^{12}\!-\!8 z^{14}\!+\!8 z^{16}\!-\!5 z^{18}\!+\!3 z^{20}\!-\!z^{22}\right)},$$
%
%$$B_2(z) = \frac{2 z^6 \left(1\!-\!z^2\right) \left(4\!-\!4 z^2\!-\!8 z^4\!+\!8 z^6\!-\!4 z^8\!+\!16 z^{10}\!-\!12 z^{12}\!+\!18 z^{14}\!-\!10 z^{16}\!+\!3 z^{18}\!-\!3 z^{20}\!-\!4 z^{22}\!+\!10 z^{24}\!-\!10 z^{26}\!+\!5 z^{28}\!-\!z^{30}\right)}{\left(1\!-\!z^2\!-\!z^4\!+\!z^6\!-\!z^8\right)^2 \left(1\!-\!3 z^2\!+\!3 z^4\!-\!5 z^6\!+\!8 z^8\!-\!9 z^{10}\!+\!7 z^{12}\!-\!8 z^{14}\!+\!8 z^{16}\!-\!5 z^{18}\!+\!3 z^{20}\!-\!z^{22}\right)}.$$}
%\normalsize
%
%{\bf Can someone who knows more Latex than me make $A_2$ and $B_2$ look more presentable please?}
These generating functions have simple poles, the dominant pole being at $z=z_c(T)>z_c(T+1) > z_c.$ 
Furthermore, $\lim_{T \to \infty} z_c(T)=z_c$ \cite{JOW06}.

Duminil-Copin and Smirnov proved that the unique solution of 
\be\label{eqn:ident_generalz}
\cos(3\pi/8)A_T(z) + B_T(z) =1,
\ee
for any $T\geq 0$, occurs at $z=z_{\rm c} = 1/\sqrt{2+\sqrt{2}}$. It follows then that we could work backwards: given only the simple rational generating functions $A_0(z)$ and $B_0(z)$, we could identify the exact value of $z_{\rm c}$ simply by seeking the solution of
$$\cos(3\pi/8)A_{0}(z) + B_{0}(z) =1.$$
% for $x > 0$ of $$\cos(3\pi/8)A_{0}(x) + B_{0}(x)-\cos(3\pi/8)A_{1}(x) - B_{1}(x)=0.$$ In this way we find, from the above equations, $x = x_c = 1/\sqrt{2 + \sqrt{2}}.$ 
If we did not already know $z_c,$ this would be a particularly simple way to find it.

In a further demonstration of this invariant, we show in Figure~\ref{hcAB} a plot of\\$\cos(3\pi/8)A_{T}(z)~+~B_{T}(z)$ for $T \in [1\ldots10],$ 
where it can be seen that the curves intersect at $(z_c,1),$ in accordance with the identity (\ref{keyh}).

Let us assume that we didn't even know the Duminil-Copin and Smirnov identity (\ref{keyh}), but rather just 
conjectured that some linear combination of $A_{T}(z_c)$ and $B_{T}(z_c)$ was invariant. We write this 
invariant as $\lambda A_{T}(z_c) + B_{T}(z_c).$ Then by seeking the solutions, for $z > 0,$ $\lambda > 0$ of the equations
$$\lambda A_{0}(z) + B_{0}(z)-\lambda A_{1}(z) - B_{1}(z)=0,$$ and
$$\lambda A_{2}(z) + B_{2}(z)-\lambda A_{1}(z) - B_{1}(z)=0,$$ 
we could discover both the invariant and the exact value of the critical point from the exact solutions for strips 
of width 0, 1 and 2 given above. 

As we show below, this suggest a way to {\em approximate} $z_c$ for other lattices, by similar means. 
We first show that, for other lattices, an appropriate linear combination of $A_T(z)$ and $B_T(z)$ 
approaches a limit as $T \to \infty,$ and use this observation to estimate the critical point for the square and triangular lattices.

\begin{figure}[h]
\begin{center}
\includegraphics[height=250pt]{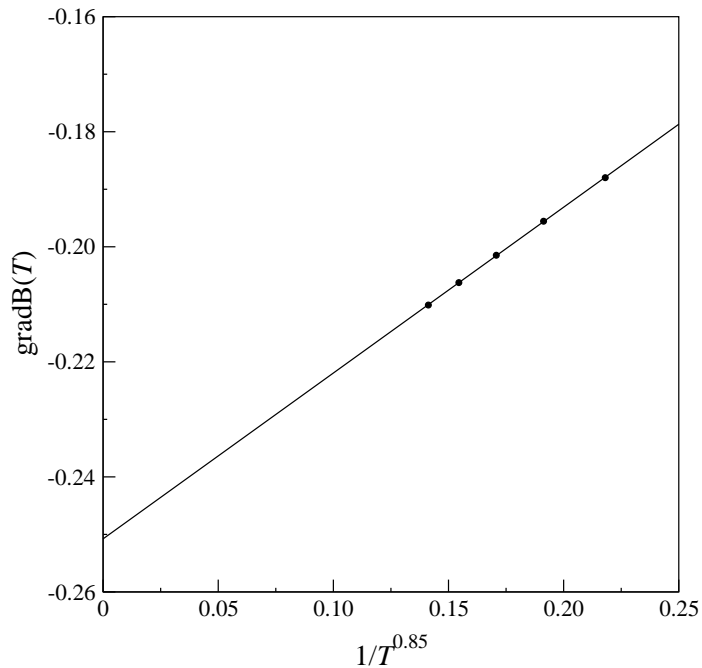}
\end{center}
\caption{Plot of the local gradient of  $B_T(z_c)$ and $T$ against $1/T^{0.85}.$ 
The straight line is a linear fit to the data in the plot.}
\label{fig:hcBgrad}
\end{figure}

Letting the width go to infinity, another of the generating functions, $B(z),$ vanishes \cite{BdGG11}, and the identity 
reduces to $A(z_c) = 1/\cos(3\pi/8).$ Recall that $A(z)$ is the generating function for loops, whose asymptotic behaviour is believed to be \cite{GT84} 
$$A(z) \sim a_0+a_1(1-z/z_c)^{3/16}.$$ 
Thus the Duminil-Copin/Smirnov identity in the limit $L \to \infty$ and $T \to \infty$ gives us the exact value 
for the amplitude term $a_0=A(z_c)$ for the honeycomb lattice.

Next we consider the behaviour of the generating functions $A_T(z)$ and $B_T(z)$ in the limit $T \to \infty.$ 
Denote $\lim_{T \to \infty} B_T(z)$ by $B(z),$ with a similar definition of $A(z).$ 
Recall that, as proved in \cite{BdGG11}, $B(z) = 0$ for $z \le z_c.$ We then wish to understand exactly
how $B_T(z_c) \to 0$ as $T \to \infty$.

If $B_T(z_c) \sim {\rm const.}/T^\alpha,$ a log-log plot of $B_T(z_c)$ against $T$ should be linear with slope 
$-\alpha$ as $T \to \infty.$ We only  have data for width $T \le 10,$ so the gradient is still changing slightly 
with $T$ in that plot. To accommodate this, we extrapolate estimates of the local gradient.
We define the local gradient as 
$${\rm gradB}(T)=\log\left(\frac{B_T(z_c)}{B_{T-1}(z_c)}\right)/\log\left(\frac{T}{T-1}\right)$$ 
and plot ${\rm gradB}(T)$ against $1/T^{0.85},$ where the exponent $0.85$ was chosen empirically to 
make the plot linear. The plot is shown in Figure \ref{fig:hcBgrad}, and it is manifestly clear that the locus 
 extrapolates to a value of $\alpha \approx 1/4.$ More systematic numerical extrapolation techniques \cite{G89} 
(not detailed here) lend support to this estimate. This is precisely as predicted  \cite{DC-S10}, based on
conjectures of Lawler, Schramm and Werner  \cite{LSW04} as to the number of SAWs on the boundary of a domain. 
This is discussed in \cite{DC-S10}, where it is pointed out that the conjecture implies that $B_T(z_c)$ should decay 
as $T^{-1/4}$ as $T$ goes to infinity, just as we observe.

From (\ref{keyh}) it follows that if $B_T(z_c) \sim c/T^{1/4},$ then  $\tilde {A}_T(z_c) =  A_T(z_c) - a_0 $ also decays 
as  $T^{-1/4},$ and this was observed numerically by a similar plot to that described in the preceding paragraph.

\section{Square lattice}\label{square}
As discussed above, there is no parafermionic operator that applies to the SAW model 
on the square lattice or the triangular lattice, so we can't  identify the critical point for SAW on these 
lattices as did Duminil-Copin and Smirnov for honeycomb SAW. We might, however, expect that in the 
limit $L \to \infty$ (so that we are again considering  SAW in a strip) there should be a similar relationship 
between the two generating functions $A_T(z)$ and $B_T(z),$ with some $T$-dependence that vanishes as $T \to \infty.$

That is to say, while the relationship 
$$1=\cos(3\pi/8)A_{T}(z_c) + B_{T}(z_c),$$ 
which is an identity for honeycomb lattice SAW for finite width $T$, cannot be expected to hold 
for the square and triangular lattices,  we might expect something like 
$$1=c_\alpha(T)A_T(z_c) + c_\beta(T)B_T(z_c)$$
to hold, where the ``constants" $c_\alpha(T)$ and $c_\beta(T)$ are weakly $T$-dependent.

We have computed data for the square lattice generating functions in strips of width $T,$ that is, 
$A_T(z)$ and $B_T(z),$ for $T \le 15,$ and used our best estimate $1/z_c=2.63815853031$ \cite{JG99,IJ03} 
to tabulate $A_T(z_c)$ and $B_T(z_c),$ shown in Table~\ref{tabsq}.  In \cite{DC-S10} these generating 
functions for the honeycomb lattice were defined to include an extra half-step at the beginning of the 
walk and at the end of the walk. This introduces an extra factor of $z$ (or, as appropriate $z_c,$) and 
we have used this definition of the generating functions  $A_T(z)$ and $B_T(z)$ for the  square lattice 
data.

\begin{table}
\begin{center}
\begin{tabular}{|c|c|c|} \hline 
$T$ & $A_T(z_c)$ & $B_T(z_c)$ \\ \hline
1 &    0.684928096008073 &    0.760082094484555 \\
2 &    0.825972541624066 &    0.707257323612670 \\
3 &    0.927565166390104 &    0.668934606497192 \\
4 &    1.006072923950508 &    0.639202723889591 \\
5 &    1.069537792384553 &    0.615108345881821 \\
6 &    1.122482001562161 &    0.594974760428940 \\
7 &    1.167689112421950 &    0.577763265643123 \\
8 &    1.206987841982332 &    0.562788338725227 \\
9 &    1.241640411741764 &    0.549575210877016 \\
10 &    1.272552495675558 &    0.537782341996967 \\
11 &    1.300394615482380 &    0.527156358502502 \\
12 &    1.325676196007041 &    0.517504450137522 \\
13 &    1.348792763213512 &    0.508676719252903 \\
14 &    1.370057142972426 &    0.500554481834765 \\
15 &    1.389720731591218 &    0.493042273647721 \\
\hline
\end{tabular}\\
\end{center}
\caption{The values of $A_T(z)$ and $B_T(z)$ at the critical point $z=z_c$.}  \label{tabsq}
\end{table}

\begin{figure}[h]
\begin{center}
\includegraphics[height=250pt]{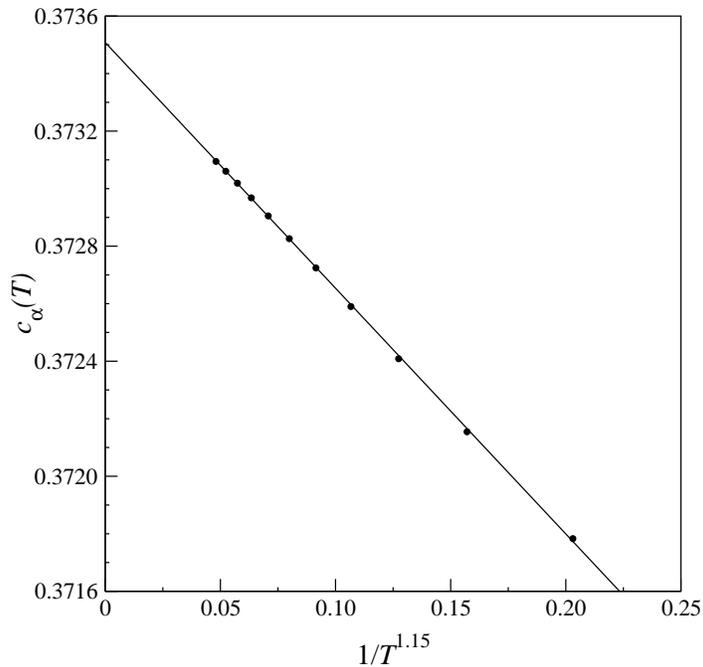}
\end{center}
\caption{Plot of $c_{\alpha}(T)$ against $1/T^{1.15}$ for square lattice $A$ walks. 
The straight line is a linear fit to the last seven data-points in the plot.}
\label{casq}
\end{figure}

\begin{figure}[h]
\begin{center}
\includegraphics[height=250pt]{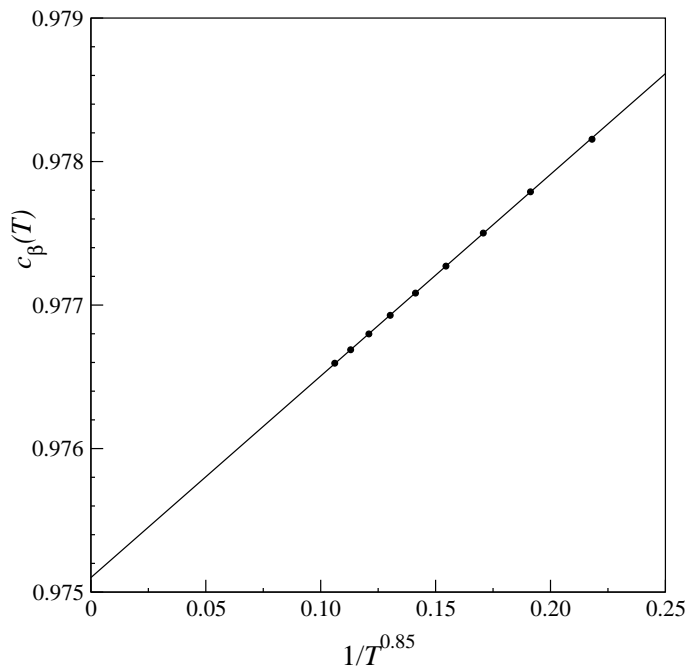}
\end{center}
\caption{Plot of $c_{\beta}(T)$ against $1/T^{0.85}$ for square lattice $B$ walks.
The straight line is a linear fit to the last seven data-points in the plot. }
\label{cbsq}
\end{figure}

We then fitted successive pairs of values $(A_T(z_c),B_T(z_c))$ and $(A_{T+1}(z_c),B_{T+1}(z_c))$ for $T=1\dots 14$ 
to  $$1 =c_\alpha(T) A_{T}(z_c) + c_\beta(T) B_{T}(z_c),$$ and solved the associated linear equations for 
$c_\alpha(T)$ and $c_\beta(T),$ using our best estimate of $z_c.$ In Figures \ref{casq} and \ref{cbsq} we show plots of values of 
$c_\alpha(T)$ against $1/T^{1.15}$ and $c_\beta(T)$ against $1/T^{0.85}.$ 

We have no basis for assuming that this is the correct form we should choose to extrapolate these plots; 
rather, the $T$-dependence was chosen experimentally to give a linear plot.  Extrapolated to $T = \infty,$ 
we find $c_\alpha \approx 0.3734 $ and $c_\beta \approx 0.9756.$ To obtain more precise estimates, 
we extrapolated these sequences using the Bulirsch-Stoer algorithm \cite{BS64}. This algorithm requires a parameter
$w$ which can be thought of as a correction-to-scaling exponent. For the purpose of the current exercise, we have set this parameter to $1$, corresponding to an analytic correction, which is appropriate for the two-dimensional SAW problem \cite{CGJ05}. Our implementation of the algorithm is precisely as described by Monroe \cite{M02}, and we retained 40 digit precision throughout.
We also applied a range of standard extrapolation algorithms to the sequences $\{ c_\alpha(T) \}$ and $\{ c_\beta(T) \}.$ 
These were Levin's $u$-transform, Brezinskii's $\theta$ algorithm, Neville tables, Wynn's $\epsilon$ algorithm 
and the Barber-Hamer algorithm. Descriptions of these algorithms, and codes for their implementation, 
can be found in \cite{G89}. These gave results totally consistent with, but less precise than, those from the Bulirsch-Stoer algorithm.

In this way we estimated $c_\alpha = 0.373362 \pm 0.000001 $ and $c_\beta = 0.975644 \pm 0.000002.$ Thus the ratio 
$c_\alpha/c_\beta = 0.382683 (2).$   For the honeycomb lattice the corresponding ratio is 
$\cos(3\pi/8) = 0.3826834\ldots,$ which is close to, and probably equal to, the square lattice value. 
We shall see in the next section that this apparent agreement also holds for the triangular lattice.

Assuming that $c_\alpha/c_\beta = \cos(3\pi/8)$ for the square lattice, we calculated elements of the 
sequence $\cos(3\pi/8)A_T(z_c)+B_T(z_c)$ and extrapolated these using the same  extrapolation 
method as described above. We found the limit of the sequence to be $1.024966 \pm 0.000001,$ compared 
to a value of exactly $1$ for the honeycomb lattice. Using this estimate  we plotted  
$\log({\cos(3\pi/8)A_{T}(z_c) +  B_{T}(z_c) -1.024966})$ against $\log{T}.$  The plot displayed slight 
curvature, so we plotted the local gradient,
\begin{equation*}
\log\left(\frac{\cos(3\pi/8)A_{T}(z_c) +  B_{T}(z_c) -1.024966}{\cos(3\pi/8)A_{T-1}(z_c)+  B_{T-1}(z_c) -1.024966}\right)/\log\left(\frac{T}{T-1}\right)
\end{equation*}
against $1/T.$ This extrapolated to a value in the range $(1.9, 2.1)$, so we took the central value and concluded that
 $1.024966 - \frac{c_1}{T^{2}} \approx \cos(3\pi/8)A_{T}(z_c) +  B_{T}(z_c)$ is the asymptotic behaviour. Finally, extrapolating 
 estimates of the constant $c_1,$ we estimate $c_1 \approx 0.14 \pm 0.02.$ So our final result is 
 $$1.024966 - \frac{0.14}{T^{2}}\approx\cos(3\pi/8)A_{T}(z_c) +  B_{T}(z_c), $$ 
 which is an accurate mnemonic for square lattice strips.

For the honeycomb lattice, it has been proved \cite{BdGG11} that $\lim_{T \to \infty} B_T(z_c) = B(z_c) = 0.$ 
The proof applies {\em mutatis mutandis} to the square and triangular lattices. Thus in the limit of infinite 
strip width we find $1\approx 0.3733621A(z_c),$ giving a prediction for the critical amplitude 
$A(z_c) \approx 2.678365.$ Current series estimates (unpublished) are $2.66\pm0.03$, 
some 4 orders of magnitude less accurate than this new estimate.

For the honeycomb lattice the intersection point of $\cos(3\pi/8)A_{T}(z) + B_{T}(z)$ for any two distinct 
values of $T$ uniquely determines $z_c.$ For the square and triangular lattices, we instead looked at the 
intersection point of $\cos(3\pi/8)A_{T}(z) + B_{T}(z)$ and $\cos(3\pi/8)A_{T+1}(z) + B_{T+1}(z).$ 
Call this intersection point $z_c(T).$ Then one expects $\lim_{T \to \infty} z_c(T) = z_c.$ We extrapolated the sequence
$\{z_c(T)\}$ using the same Bulirsch-Stoer method described above, and in this way we estimated 
$$z_c = 0.3790522775 \pm 0.0000000005.$$ 
This estimate can be compared to the best series estimates, 
based on analysis of very long polygon series, $z_c({\rm sq})=0.37905227776$  \cite{JG99,IJ03},  
with uncertainty in the last digit.
Thus this method is seen to be a powerful new method for estimating critical points, giving very good accuracy, 
though it doesn't rival the most powerful methods based on series analysis of polygon series.
 However it does give comparable accuracy to methods based on series analysis of SAW (rather than SAP). 

\subsection{$T$-dependence of the generating functions $A_T(z_c)$ and $B_T(z_c).$}

\begin{figure}[h!]
\begin{center}
\includegraphics[height=250pt]{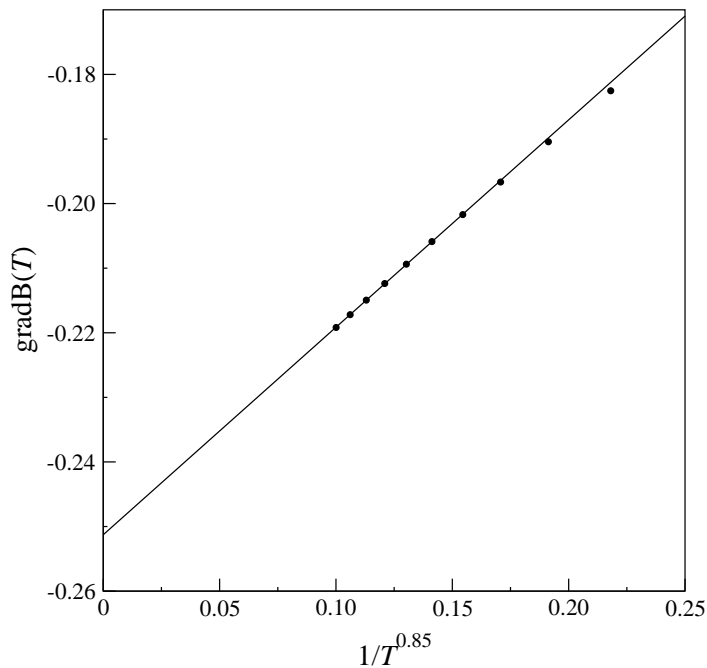}
\end{center}
\caption{Plot of ${\rm gradB}(T)$ against $1/T^{0.85}$ for square lattice $B$ walks. 
The straight line is a linear fit to the last seven data-points in the plot. Linear extrapolation to -0.25 is well supported. }
\label{sqBgrad}
\end{figure}

As for the honeycomb lattice, we expect $B_T(z_c) \sim {\rm const.}/T^{1/4}$. In Figure~\ref{sqBgrad} 
we have plotted estimates of the exponent 
$${\rm gradB}(T) =\log\left(\frac{B_T(z_c)}{B_{T-1}(z_c)}\right)/\log\left(\frac{T}{T-1}\right)$$ 
against $1/T^{0.85}$. This local gradient should approach $-1/4$ and from the plot is seen to do so.

As for the honeycomb lattice, from (\ref{keyh}) it follows that if $B_T(z_c) \sim {\rm const.}/T^{1/4},$ then 
 $\tilde {A}_T(z_c) =  A_T(z_c) - a_0 $ also decays as  $T^{-1/4}.$ This was observed numerically 
 by a similar plot to that described in the preceding paragraph.

\subsection{Alternative estimate of the critical point}
We showed in Section \ref{honey} that the critical point could be identified just from knowledge of the invariant 
(\ref{keyh}) and the formulae for $A_0(z)$ and $B_0(z).$
In the case of the square lattice, we expect the simultaneous solution of the pair of equations

$$\lambda A_{T-1}(z) + B_{T-1}(z)-\lambda A_{T}(z) - B_{T}(z)=0,$$ and
$$\lambda A_{T}(z) + B_{T}(z)-\lambda A_{T+1}(z) - B_{T+1}(z)=0,$$
to give a sequence of estimates of $z_c(T)$ that should converge to the critical point $z_c.$ Similarly, the parameter $\lambda$ should converge to the ratio $c_\alpha/c_\beta = \cos(3\pi/8).$  The merit of this method of estimating the critical point is that it makes no assumption about the value of $\lambda,$ while  the method of estimating the critical point described in Section \ref{square}
assumes that $\lambda = \cos(3\pi/8).$ 

We solved these equations by seeking the solution of 
$$(A_{T-1}(z) - A_{T}(z))(B_{T+1}(z) - B_{T}(z)) = (A_{T}(z) - A_{T+1}(z))(B_{T}(z) - B_{T-1}(z)),$$ which we call $z_c(T),$
and then found $\lambda(T)$ by back substitution. The results are shown in Table~\ref{sqxcl}. 

We plotted (not shown) the estimates of $z_c(T),$ against various powers of $1/T,$ and found 
a linear plot if we plotted against $1/T^2.$ We extrapolated the estimates $z_c(T)$ for steadily 
increasing $T$ values using the same Bulirsch-Stoer extrapolation method described above. 
Rapid convergence was observed, and we estimate 
$z_c = 0.37905228 \pm 0.00000001.$ This is consistent with the limit found from our previous method described above, though not quite as precise.

We have similarly extrapolated the estimates of $\lambda(T),$ and find $\lambda \approx 0.38268,$ compared to 
the expected value $\cos(3\pi/8) =0.382682.$\\
\newline

\begin{table}[h]
\begin{center}
\begin{tabular}{|c|c|c|} \hline
$T$ & $z_c(T)$ & $\lambda(T)$ \\ \hline
2 &  0.3792132510996564 & 0.3680865016160631 \\
3 &  0.3791354275802486 & 0.3724957043646600\\
4 &  0.3791014587212902 & 0.3750327779790171\\
5 &  0.3790837649841775 & 0.3766921607963391\\
6 &  0.3790736177161640 & 0.3778473665510655\\
7 &  0.3790673896037665 & 0.3786868392606266\\
8 &  0.3790633602596354 & 0.3793176229093515 \\
9 &  0.3790606406190476 & 0.3798046222881576 \\
10 &0.3790587398862656 & 0.3801891415440993 \\
11 &0.3790573721782793 & 0.3804985249279112\\
12 &0.3790563633515162 & 0.3807514859621669\\
13 &0.3790556032334455 & 0.3809611990869139\\
14 & 0.379055019822686  & 0.3811371681858631\\
\hline 
\end{tabular}
\end{center}
\caption{Estimates of $z_c(T)$ and $\lambda(T)$ for the square lattice.}  \label{sqxcl}
\end{table}
%\caption{Estimates of $z_c(T)$ and $\lambda(T)$ for the square lattice}

\section{Triangular lattice} \label{triang}

We have also generated data for the triangular lattice in strips of widths up to and including 11. 
Using the best estimate \cite{IJ04b} of the critical point, $z_c=0.2409175745$, we show, in Table~\ref{tabtr}, 
the values of $A_T(z_c)$ and $B_T(z_c)$ for each strip width. For the triangular lattice  there are two edges incident 
upon the origin in a strip geometry and this complicates matters. To simplify things, we start and finish our SAW 
{\em on} the boundary in the case of the triangular lattice, in order to avoid the complications that arise when including an incident edge. So the extra factor of $z_c$ included in the definition of these amplitudes 
for the square and honeycomb lattice data is not present in the triangular lattice data.

\begin{table}
\begin{center}
\begin{tabular}{|c|c|c|} \hline 
$T$ & $A_T(z_c)$ & $B_T(z_c)$ \\ \hline
1 &  1.139480549210468 &  1.457161363236105 \\
2 &  1.435344242350752 &  1.348134252648887 \\
3 &  1.641756149326264 &  1.270897362392145 \\
4 &  1.798515045521241 &  1.211810836367619 \\
5 &  1.923848231267622 &  1.164374555192450  \\                     
6 &  2.027608945103857 &  1.125001488941636 \\
7 &  2.115709764900265 &  1.091512525007183  \\
8 &  2.191966367371986 &  1.062490013670246  \\
9 &  2.258977760090717 &  1.036962918106255 \\
10 &2.318589791981952 &  1.014238779515961 \\
11 &2.372157936598986 &  0.993807536013206 \\ \hline
\end{tabular}
\end{center}
\caption{Estimates of $A_T(z_c)$ and $B_T(z_c)$ for various strip widths for the triangular lattice.} \label{tabtr}
\end{table}
                     
As in the analysis of the square lattice data, we fitted successive pairs of values $(A_T(z_c),B_T(z_c))$ 
and $(A_{T+1}(z_c),B_{T+1}(z_c))$ for $T=1\dots 10$ to  
$$1 =c_\alpha(T) A_{T}(z_c) + c_\beta(T) B_{T}(z_c),$$ 
and solved the associated linear equations for $c_\alpha(T)$ and $c_\beta(T).$ 
%In Figures \ref{casq} and \ref{cbsq} we show plots of values of $c_\alpha(T)$ against $1/T^{1.15}$ and $c_\beta(T)$ against $1/T^{0.85}.$ 

%We have no basis for assuming that this is the correct form we should choose to extrapolate these plots, rather the $T$ dependence was chosen experimentally to give a linear plot.  Extrapolated to $T = \infty,$ we find $c_\alpha \approx 0.1415 $ and $c_\beta \approx 0.3698.$ 

To obtain  precise estimates, we again applied the Bulirsch-Stoer extrapolation algorithm
to the sequences $\{ c_\alpha(T) \}$ and $\{ c_\beta(T) \}.$  Combining the results from these different algorithms, we estimate 
$c_\alpha = 0.2012028 (3) $ and $c_\beta = 0.525770 (3).$ Thus the ratio $c_\alpha/c_\beta = 0.382682 (3).$  
For the honeycomb lattice the corresponding ratio is $\cos(3\pi/8) = 0.3826834\ldots,$ which (as we also saw 
for the square lattice) is close to, and probably equal to, the triangular lattice value. 

Assuming the ratio $c_\alpha/c_\beta = \cos(3\pi/8)$ for the triangular lattice too,  we extrapolated 
$\cos(3\pi/8)A_{T}(z_c) + B_{T}(z_c)$ for increasing values of $T,$ using our standard suite of 
extrapolation algorithms and the Bulirsch-Stoer algorithm. We estimated the limit to be $1.901979 \pm 0.000001.$ We then repeated
the analysis described above for the square lattice data {\em mutatis mutandis}, and found
$$1.901979 - \frac{0.1}{T^{2}} \approx \cos(3\pi/8)A_{T}(z_c) +  B_{T}(z_c). $$
As remarked above, it has been proved \cite{BdGG11} that $\lim_{T \to \infty} B_T(z_c) = B(z_c) = 0.$  
Thus in the limit of infinite strip width we find $1.901979 \approx\cos(3\pi/8)A(z_c),$ a prediction for 
the critical amplitude $A(z_c) \approx 4.970111.$

 As for the square lattice case, we estimated the critical point $z_c$ by extrapolating the 
intersection point of $\cos(3\pi/8)A_{T}(z) + B_{T}(z)$ and $\cos(3\pi/8)A_{T+1}(z) + B_{T+1}(z),$ called $z_c(T).$ One expects $\lim_{T \to \infty} z_c(T) = z_c.$ 

In this way we estimated 
$$z_c = 0.240917572 \pm 0.000000005$$ 
for the triangular lattice.  This can be compared to the best series estimate, 
based on analysis of very long polygon series $z_c({\rm tr}) = 0.2409175745$  \cite{IJ04b}
with uncertainty in the last quoted digit.

\subsection{$T$-dependence of the generating functions $A_T(z_c)$ and $B_T(z_c).$}
                            
As for the honeycomb and square lattices, we expect $B_T(z_c) \sim {\rm const.}/T^{1/4}$. 
We plotted estimates of the exponent 
$${\rm gradB}(T) =\log\left(\frac{B_T(z_c)}{B_{T-1}(z_c)}\right)/\log\left(\frac{T}{T-1}\right)$$ 
against $1/T^{0.85},$ which should approach $-1/4$, and were seen to do so. The figure was 
visually indistinguishable from the corresponding Figure~\ref{sqBgrad} for the square lattice, so is not shown.

 Similarly, it follows from (\ref{keyh}) that $\tilde {A}_T(z_c) =  A_T(z_c) - a_0 \approx A_T(z_c) - 4.97011 $ 
 decays as $1/T^{1/4}$  as $T$ tends to infinity. As we did for the square lattice case, we also confirmed this numerically.
 
 \subsection{Alternative estimate of the critical point}

In the previous section, we showed that, for the square lattice data, the simultaneous solution of the pair of equations

$$\lambda A_{T-1}(z) + B_{T-1}(z)-\lambda A_{T}(z) - B_{T}(z)=0,$$ and
$$\lambda A_{T}(z) + B_{T}(z)-\lambda A_{T+1}(z) - B_{T+1}(z)=0,$$
gives a sequence of estimates of $z_c(T)$ that  converges to the critical point $z_c.$ Similarly, 
the parameter $\lambda$  converges to the ratio $c_\alpha/c_\beta = \cos(3\pi/8),$ where the 
equality is conjectural. We solved these equations using the triangular lattice data, by seeking the solution of 
$$(A_{T-1}(z) - A_{T}(z))(B_{T+1}(z) - B_{T}(z)) = (A_{T}(z) - A_{T+1}(z))(B_{T}(z) - B_{T-1}(z)),$$ called $Z_c(T),$
and then found $\lambda(T)$ by back substitution. The results are shown in Table~\ref{trxcl}. 

We plotted (not shown) the estimates of $z_c(T),$ against various powers of $1/T,$ and found a 
linear plot if we plotted against $1/T^2.$ We analysed the sequences in precisely the same way 
as for the corresponding square lattice data, using the Bulirsch-Stoer algorithm. The merit of this method of estimating the critical point is that it makes no assumption about the value of $\lambda,$ while  the method of estimating the critical point described in Section \ref{triang}
assumes that $\lambda = \cos(3\pi/8).$  For the 
critical point we estimate $z_c = 0.240917575 \pm 0.000000005.$ This is of comparable precision to our estimate given in Section \ref{triang}, but slightly less precise than the 
best series estimate \cite{IJ04b} of $z_c=0.2409175745,$ 
with uncertainty in the last digit. 

Thus  this method is again seen to be a powerful one for estimating critical points, giving very good 
accuracy.  We 
have similarly extrapolated the estimates of $\lambda(T),$ and find $\lambda \approx 0.38268,$ 
compared to the expected value $\cos(3\pi/8)=0.382682,$ exactly as for the square lattice.\\
\newline

\begin{table}[h]
\begin{center}
\begin{tabular}{|c|c|c|} \hline
$T$ & $z_c(T)$ & $\lambda(T)$ \\ \hline
2   &  0.241168440165255 & 0.356318356471223 \\
3   &  0.241030169141752 & 0.366143831978748\\
4   &  0.240977832351101 & 0.371147184391665\\
5   &  0.240953612190006 & 0.374091365359823\\
6   &  0.240940839933527 & 0.375992279128027\\
7   &  0.240933460889859 & 0.377300697288292\\
8   &  0.240928899289076 & 0.378244599187661\\
9   &  0.240925927855959 & 0.378950553554884\\
10 &  0.240923909640445 & 0.379493901730187\\
\hline 
\end{tabular}
\end{center}
\caption{Estimates of $z_c(T)$ and $\lambda(T)$ for the triangular lattice.}  \label{trxcl}
\end{table}
\section{Enumeration of self-avoiding walks \label{sec:flm}}

The algorithm we use to enumerate SAWs on the square lattice builds on the 
pioneering work of Enting \cite{IGE80e} who enumerated square lattice 
self-avoiding polygons using the finite lattice method. More specifically 
our algorithm is based in large part on the one devised by Conway, Enting and 
Guttmann \cite{CEG93} for the enumeration of SAWs. The details of our
algorithm can be found in \cite{IJ04}.    Below we shall only briefly
outline the basics of the algorithm and describe the changes made
for the particular problem studied in this work.

\begin{figure}
\begin{center}
\includegraphics[scale=0.7]{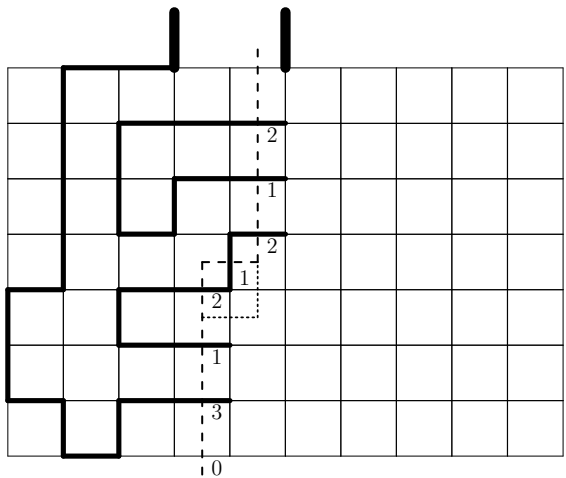}
\end{center}
\caption{\label{fig:transfer}
A snapshot of the boundary line (dashed line) during the transfer matrix 
calculation of type $A$ configurations on a  strip of size $7\times 10$. 
SAWs are enumerated by successive moves of the kink in the boundary line, as 
exemplified by the position given by the dotted line, so that one vertex and two 
edges at a time are added to the strip. To the left of the boundary line we have drawn 
an example of a partially completed SAW. The heavy lines at the top are the
incoming and outgoing edges of the SAW.}
\end{figure}

The generating function for a rectangle was calculated using transfer matrix (TM) techniques. 
The most efficient implementation of the TM algorithm generally involves 
bisecting the finite lattice with a boundary (this is just a line in the 
case of rectangles) and moving the boundary in such a way as to build up 
the lattice vertex by vertex as illustrated in Figure~\ref{fig:transfer}.  If we draw a SAW and 
then cut it by a line we observe that the partial SAW to the left of this 
line consists of a number of loops connecting two edges (we shall refer to 
these as loop ends) in the intersection, and pieces which are connected to 
only one edge (we call these free ends). The other end of the free piece is 
an end point of the SAW so there are at most two free ends. 

 Each end of a loop is assigned one of two labels 
depending on whether it is the lower end or the upper end of a loop. Each 
configuration along the boundary line can thus be represented by a set of 
edge states $\{\sigma_i\}$, where

\begin{equation}\label{eq:states}
\sigma_i  = \left\{ \begin{array}{rl}
0 &\;\;\; \mbox{empty edge},  \\ 
1 &\;\;\; \mbox{lower loop-end}, \\
2 &\;\;\; \mbox{upper loop-end}. \\
3 &\;\;\; \mbox{free end}. \\
\end{array} \right.
\end{equation}
\noindent
If we read from the bottom to the top, the configuration or signature $S$ along the 
intersection of the partial SAW in Figure~\ref{fig:transfer} is $S=\{031212120\}$. 
Since crossings aren't permitted this encoding uniquely describes 
which loop ends are connected.

A few changes to the algorithm described in \cite{IJ04} are required in order to
enumerate the restricted SAWs we study here. Most importantly the SAW must
have a free end at the middle vertex of the top side of the strip. This is easily ensured by
restricting the updating rules at this vertex (also signatures prior to passing
this vertex can have at most one free end). Specifically the middle vertex is
reached when the TM boundary has been moved halfway through the strip.
At this point the incoming edge to the left of the middle vertex is either empty,
an upper loop-end or free. In the {\it empty} case we have to insert a new free
end (along either the horizontal or vertical outgoing edge). In the {\it upper}
case the loop-end is terminated and the matching lower loop-end becomes
a free end. 

While in the {\it free} case the end is again terminated and all
the edges connected to this free end form a SAW. However this is only
a valid configuration if all other edges are empty since otherwise we 
would form configurations with more than one component.  Secondly,
in enumerating SAWs of type $A$ the second free end must lie in the
top side of the rectangle; we chose to force the free end to lie to the left of the 
middle vertex and use symmetry to count all possible configurations. In counting
bridges or SAWs of type $B$ the second free end must lie at the bottom of the strip.
Thirdly, in \cite{IJ04} the SAWs were forced to span the rectangle, that is touch
all sides, but this restriction is lifted in this study.

The sum over all contributing graphs is calculated as the boundary 
is moved through the lattice.  For each configuration of occupied or empty edges 
along the intersection we maintain a generating function $G_S$ for partial walks 
with signature $S$. In exact enumeration studies  $G_S$ would be a 
truncated two-variable polynomial  $G_S(z)$ where $z$ is conjugate to the
number of steps. 

 In a TM update each source 
signature $S$ (before the boundary is moved) gives rise to a few new target signatures 
$S'$ (after the move of the boundary line) and $n=0, 1$ or 2 new edges 
are inserted leading to the update  $G_{S'}(z)=G_{S'}(z)+z^nG_S(z)$.
Once a signature $S$ has been processed it can be discarded. 
In most studies the calculations were done using integer arithmetic  
modulo several primes with the full integer coefficients reconstructed at
the end using the Chinese remainder theorem. Here we are not really 
interested in the exact coefficients. This makes life a little easier for us
since we can use real coefficients  with the generating functions truncated at some
 maximal degree $M$. The calculations were carried out using quadruple (or 128-bit) 
 floating-point precision (achieved in FORTRAN with the REAL(KIND=16) type declaration).
 
 In our calculations we truncated $A_T(z)$ and $B_T(z)$ at degree $M=1000$ and used strips
 of half-length $L=M$. These choices of $M$ and $L$ more than suffice to ensure that
 numerical errors are negligible as evidenced by the fact that when we solve~\eqref{eqn:li3} 
 (with $z_c$ replaced by $z$) to find $z_c$ for the honeycomb lattice the estimate for
 $z_c$ agrees with the exact value to at least 30 digits, that is, to within the numerical
 accuracy of the floating-point computation itself.
 
 The computational complexity  of the  calculation  required to obtain the number 
 of walks of in a strip of width $T$  and length $L$ can be easily estimated. 
 Time (and memory) requirements are  basically proportional to a polynomial 
 in $M$ and $L$  times the maximal number of  signatures, $N_{\rm Conf}$, generated 
 during the calculation.  It is well established \cite{IJ03} that $N_{\rm Conf} \propto 3^T$ so
 the algorithm has exponential computational complexity.

% In Table~\ref{tab:ztest} we have listed estimates for 
% obtained from strips of width 10 and 9 (the crossing between  $A_{10}(z)$ and $A_9(z)$)
%for various values of  $M$ and $L$. Clearly the choice $M=L=1000$ suffices to estimate
%$y_c$  to more than 10 digits accuracy.
%
% \begin{table}[htdp]
%\caption{ \label{tab:ztest} The estimated value of $y_c$ from the crossing
%between $A_{10}(z_{\rm c},y)$ and $A_9(z_{\rm c},y)$  truncated at degree $M$ 
%and using strips of half-length from $M$ up to $10M$.}
%\begin{center}
%\begin{tabular}{ccccc}
%\hline
%$M$ & $L=M$ & $L=2M$ & $L=5M$ & $L=10M$ \\ \hline
%100 & 1.832547814756 & 1.778376701255 & 1.778024722094 & 1.778024722094 \\
%250 & 1.776250937231 & 1.775990603337 & 1.775990594686 & 1.775990594686 \\
%500 & 1.775990340341 & 1.775990291271 & 1.775990291271 & \\
%1000 &1.775990291271 & & & \\
%\hline
%\end{tabular}
%\end{center}
%\label{default}
%\end{table}%

 The transfer-matrix algorithm is eminently suited to parallel
computations and here we used the approach first
described in  \cite{IJ03} and refer the interested reader to this 
publication for further detail.  The bulk of the calculations for this paper
were performed on the cluster of the NCI National Facility,
which provides a peak computing facility to researchers in Australia. 
The NCI peak facility is a Sun Constellation Cluster
with 1492 nodes in Sun X6275 blades, each containing
two quad-core 2.93GHz Intel Nehalem CPUs with most nodes
having 3GB of memory per core (24GB per node). It took a total
of about 1800 CPU hours to calculate $A_T(z)$ for $T$ up to 15.
So,  the bulk of the time (almost 1250 hours) was spent calculating
$A_{15}(z)$. In this case we used 48 processors and the split between 
actual calculations and communications was roughly 2 to 1 (with
quite a bit of variation from processor to processor). Smaller widths
can be done more efficiently in that communication needs are fewer
and hence not as much time is used for this task.

 On a technical issue we note that quad precision is not a supported
 data type in the MPI standard. So in order to pass messages containing
 the generating functions we used the MPI data type MPI-BYTE with
 each coefficient then having a length of 16 bytes.

 The algorithm used for the triangular lattice is quite similar. The triangular lattice
 is represented as a square lattice with additional edges along one of the main
 diagonals. This poses an immediate problem since a boundary line drawn as
 in Figure~\ref{fig:transfer} would intersect $2T$ edges thus greatly increasing the number
 of possible  signatures. In this case it is more efficient to draw the boundary  line
 through the vertices of the lattice. We then again have $T$ intersections, however
 a vertex may be in an additional state since a partial SAW can touch the boundary
 line without crossing it (see \cite{IJ04b} for further details). The upshot is that the computational 
 complexity grows exponentially as $4^T$.

\section{Conclusion}
We started this study in order to consider to what extent, in some ill-defined sense, does there exist a 
parafermionic operator applicable to SAWs on the square and triangular lattice. 
For honeycomb lattice SAWs,~\eqref{keyh} is an identity. For the square and triangular lattices we found, 
experimentally, that for walks in a strip of width $T$ on the square lattice,
\begin{equation} \label{eq:s}
1.024966(1 - 0.14/T^2)\approx\cos(3\pi/8) A_{T}(z_{\rm c}) +  B_{T}(z_{\rm c}),
\end{equation}
  while for the triangular lattice the corresponding result is
  \begin{equation} \label{eq:t}
1.901979(1 - 0.1/T^2)\approx \cos(3\pi/8) A_{T}(z_{\rm c}) +  B_{T}(z_{\rm c}).
\end{equation}
Since $B_{T}(z_{\rm c}) \to 0$ as $T \to \infty,$ it follows that in the same limit $A_{T}(z_{\rm c}) \to a_0,$ 
where the numerical value of $a_0 = 1/\cos(3\pi/8)$ for the honeycomb lattice, $a_0 \approx 1.024966/\cos(3\pi/8)$ 
for the square lattice, and $a_0 \approx 1.901979/\cos(3\pi/8)$ for the triangular lattice.
We provided numerical support for the conjecture that $B_{T}(z_{\rm c}) \sim {\rm const.}/T^{1/4}$ as $T \to \infty,$ 
and hence that $A_{T}(z_{\rm c}) - a_0 \sim {\rm const.}/T^{1/4}$ also.
Finally we show how the existence of equations (\ref{keyh}), (\ref{eq:s}) and  (\ref{eq:t}) suggests a powerful 
numerical method to estimate the critical point. In that way we found $z_{\rm c}$ exactly for the honeycomb lattice, 
and estimated $z_{\rm c} = 0.3790522775(5) $ for the square lattice and $z_{\rm c} = 0.240917573(5) $ for the triangular lattice.

It has been pointed out to the authors by John Cardy (private communication) that in the scaling limit, all two dimensional SAW models are given by the same conformal field theory (CFT). Since it is known that for one of these models (i.e. honeycomb lattice SAW) the critical point can be found by requiring certain contour integrals to vanish (i.e. when summing~\eqref{eqn:local_identity} over a region of the lattice), it follows that in the scaling limit the same must be true for all two dimensional SAW. This is entirely consistent with our observations and the relations~\eqref{eq:s} and~\eqref{eq:t}.

  \section{Acknowledgments}
We would like to thank Jan de Gier and John Cardy for many discussions that aided our understanding.
This work was supported by an award under the Merit Allocation Scheme 
on the NCI National Facility at the ANU
 and was supported under the Australian Research Council's Discovery Projects funding scheme in grants to AJG and IJ. 
NRB was supported by the ARC Centre of Excellence for Mathematics and Statistics of Complex Systems (MASCOS).

\end{document}